\documentclass[letterpaper]{article} 
\usepackage{aaai2027}
\usepackage[hyphens]{url}  
\usepackage{graphicx} 
\usepackage{natbib}  
\usepackage{caption} 
\usepackage{algorithm}
\usepackage{algorithmic}
\usepackage{multirow}
\usepackage{makecell}
\usepackage[table]{xcolor}
\usepackage{booktabs}
\usepackage{amsmath}
\usepackage{url}

\usepackage{newfloat}
\usepackage{listings}
\DeclareCaptionStyle{ruled}{labelfont=normalfont,labelsep=colon,strut=off} 
\floatstyle{ruled}
\newfloat{listing}{tb}{lst}{}
\floatname{listing}{Listing}

\usepackage{booktabs}

\title{DAT: Dual-Aware Adaptive Transmission with Small-Large Model Collaboration for Edge-Cloud Anomaly Detection}

\author{
    Qi Guo\textsuperscript{\rm 1,\rm 3},
    Zheming Yang\textsuperscript{\rm 1}\corresponding,
    Yunqing Hu\textsuperscript{\rm 1,\rm 3},
    Chang Zhao\textsuperscript{\rm 1,\rm 3},
    Wen Ji\textsuperscript{\rm 1,\rm 2}\corresponding
}

\affiliations{
    \textsuperscript{\rm 1}State Key Laboratory of Processors,
    Institute of Computing Technology,
    CAS,
    Beijing 100190, China\\
    \textsuperscript{\rm 2}Institute of AI for Industries,
    CAS,
    Nanjing 211135, China\\
    \textsuperscript{\rm 3}University of Chinese Academy of Sciences,
    Beijing 100049, China\\
    guoqi23p@ict.ac.cn,
    yangzheming@ict.ac.cn,
    huyunqing24s@ict.ac.cn,
    zhaochang23s@ict.ac.cn,
    jiwen@ict.ac.cn
}

\begin{document}

\maketitle

\begin{abstract}
Continuous video anomaly detection is critical for timely safety monitoring in edge--cloud visual systems, but it remains challenging because abnormal events are sparse, transient, and require both low-latency anomaly detection and verifiable visual evidence. Multimodal large language models (MLLMs) offer strong semantic understanding and visual reasoning capabilities for such event-centric analysis, yet applying them to continuous video streams incurs prohibitive computation and communication overhead under bandwidth-constrained edge--cloud deployments. To address this challenge, we propose DAT, a dual-aware adaptive transmission framework that supports high-quality semantic generation, low-latency anomaly detection, and progressive visual evidence supplementation. DAT first uses a collaborative small--large model cascade, where a lightweight edge-side model filters non-target-event frames, performs object detection, and triggers MLLM inference only for suspicious frames. Building on this cascade, DAT introduces visual guidance and semantic prompting to improve structured event understanding, localization, and output consistency. It further schedules semantic anomaly alarms and visual evidence through a semantic- and bandwidth-aware multi-stream transmission optimization method. Experimental results show that DAT achieves 97.77\% F1 in traffic-accident detection while substantially reducing inference and communication overhead. On continuous video workloads, DAT lowers the MLLM call rate to 13.71\% and reduces upload volume by at least 95.72\% compared with representative edge--cloud baselines. Under severe congestion, DAT reduces weighted alarm delay by up to 87.9\%, demonstrating efficient MLLM-based anomaly detection under constrained edge--cloud networks.
\end{abstract}


\section{Introduction}

Visual anomaly detection is widely used in continuously monitored environments such as traffic surveillance and industrial inspection~\cite{zhu2024videoanomalydetection,TrafficAnomaly2023}. In these settings, anomalous events are typically infrequent, sudden, and potentially consequential, requiring timely anomaly detection~\cite{anomaly_detection2024}. Practical anomaly detection systems should not only determine whether an anomaly has occurred but also provide structured descriptions of its type, severity, and location, together with visual evidence for subsequent verification and response~\cite{WhatWhyHow2026}. Conventional visual detectors mainly produce predefined labels, bounding boxes, or object trajectories, limiting their ability to understand and describe complex anomalous events~\cite{buzumllm2025}. MLLMs~\cite{2025openaigpt5,BLIP-2023,2025qwen2.5vl} offer strong visual understanding and semantic generation capabilities, making them well suited to event-centric anomaly analysis~\cite{45,47,49,50}. However, applying MLLMs directly to continuous video streams incurs substantial inference and communication challenges as visual intelligence systems scale toward continuous deployment. Connected IoT devices are projected to reach 39 billion by 2030~\cite{IoTAnalytics2030}, while video already accounts for approximately 75\% of mobile data traffic~\cite{videostream76}.

\begin{figure}[t]
  \centering
  \includegraphics[width=0.9\columnwidth]{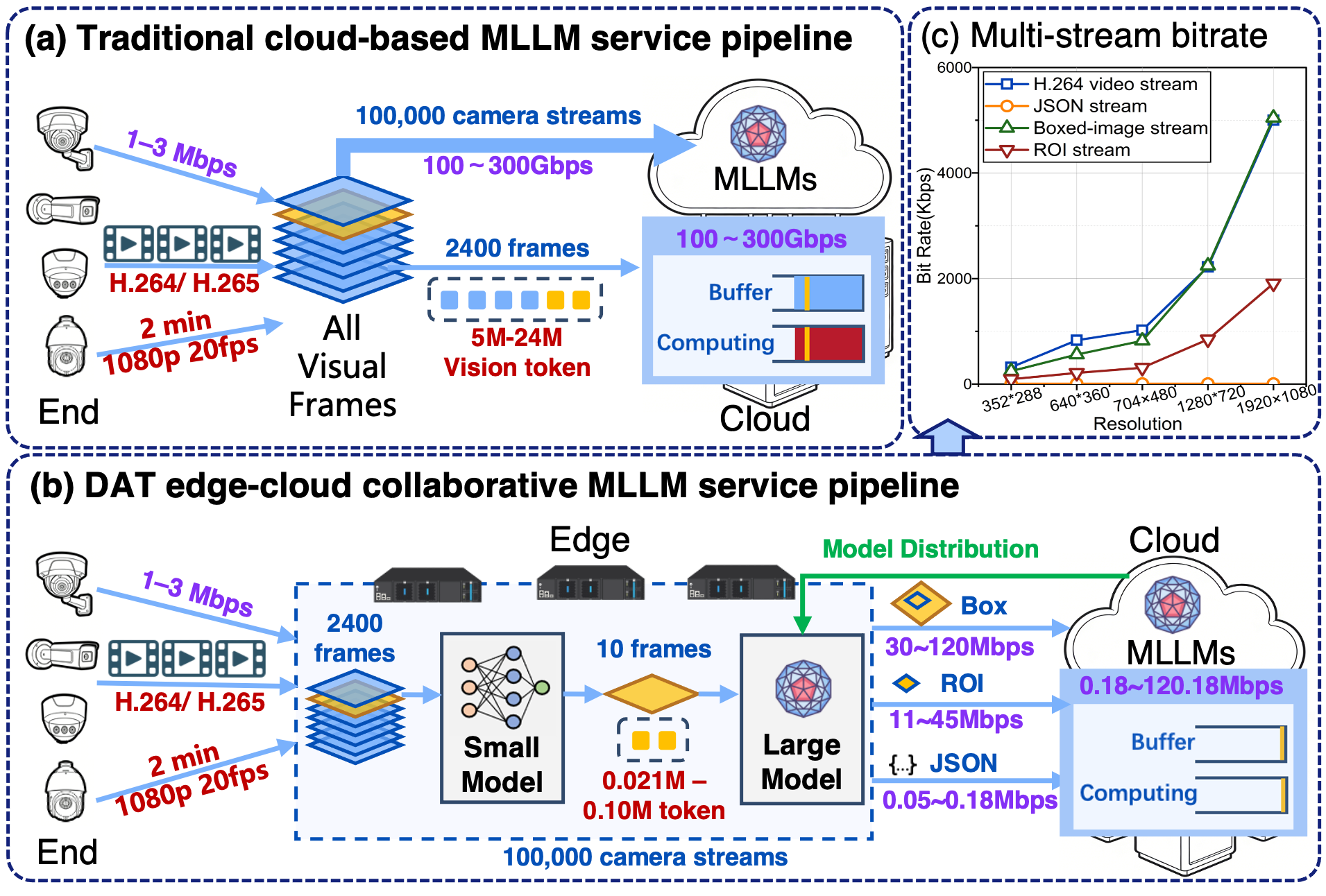}
  \caption{Comparison of traditional and DAT service pipelines.}
  \label{fig:1a}
\end{figure}

The first challenge is the high MLLM inference load caused by redundant continuous anomaly video. As shown in Figure~\ref{fig:1a}(a), a two-minute 1080p video at 20 fps contains 2,400 frames. Uniformly submitting such streams to cloud MLLMs converts many background frames into redundant visual tokens and repeated large-model calls~\cite{survey_mllm2025,VideoLLM-MoD}. The second challenge is the substantial pressure that continuous high-resolution video places on bandwidth-constrained uplinks. Even with H.264/H.265 compression, 100,000 concurrent 1080p  streams would require about 100--300 Gbps uplink bandwidth~\cite{h264/h265,5g_imt2020}. More importantly, under limited bandwidth, high-volume visual data may queue ahead of compact, time-critical semantic anomaly alarms, delaying their delivery and associated visual evidence~\cite{44,48}.


To address these two coupled challenges, we propose DAT, a Dual-Aware Adaptive Transmission with Small--Large Model Collaboration for edge--cloud anomaly detection, as shown in Figure~\ref{fig:1a}(b). DAT converts dense continuous video into sparse event-level semantic information and on-demand visual evidence, reducing both MLLM inference load and uplink transmission pressure. The main contributions of this paper are as follows:

\begin{itemize}
    \item We propose DAT, an edge--cloud anomaly detection framework that couples edge-side anomaly gating, spatial-prior-guided MLLM reasoning, and dual-aware multi-stream transmission. DAT transforms continuous visual inputs into time-critical structured semantic streams and complementary visual evidence streams, supporting timely anomaly detection under bandwidth constraints.

    \item We formulate Dual-Aware transmission as a lexicographic optimization problem and develop an online hierarchical greedy scheduler that prioritizes weighted semantic-anomaly-alarm completion time before deadline-valid visual evidence delivery. 

    \item Extensive experiments demonstrate that DAT achieves 97.77\% F1 in traffic-accident detection. On continuous video workloads, it lowers the MLLM call rate to 13.71\% and reduces upload volume by at least 95.72\% compared with the baselines. Under severe congestion, DAT reduces weighted alarm delay by up to 87.9\% while delivering 98.33\% of visual evidence within 0.5 s.
\end{itemize}

\section{Related Work}
Edge--cloud collaborative inference has become a common design choice for visual systems~\cite{46}. AppealNet~\cite{li2021AppealNet} forwards inputs that cannot be reliably handled at the edge to a stronger cloud model, while Sniper~\cite{Sniper2022} and JAVP~\cite{JAVP2023} adapt inference nodes or model configurations according to network status, device status, and task settings. Recent MLLMs, including LLaVA~\cite{LLaVA2023}, Qwen2.5-VL~\cite{2025qwen2.5vl}, and GPT-5~\cite{2025openaigpt5}, have made structured visual reasoning and semantic generation more practical~\cite{largeaimodelsemetic2024}. Building on this capability, AIVD~\cite{AIVD} uses edge detectors to produce localization candidates before cloud-side MLLM reasoning; SAEC~\cite{SAEC} improves industrial visual detection through scene-complexity awareness and adaptive edge--cloud routing; and EdgeCloudAI~\cite{EdgeCloudAI2024} reduces unnecessary cloud-side MLLM calls through edge-side filtering. Overall, these studies mainly optimize the inference path itself, including model placement and task routing. Less attention has been paid to how structured semantic information and visual evidence produced after inference should be jointly transmitted under dynamic networks.

Transmission optimization provides a foundational capability, yet its objectives are not fully aligned with MLLM-driven visual anomaly analysis. Conventional adaptive video transmission, built upon DASH/HLS-style streaming~\cite{DASH2011,apple_hls_2025}, has been extensively studied through heuristics based on throughput or buffer states~\cite{tuntu2012,buffer-based2014,buffer-based2020}, QoE or control-theoretic models~\cite{BOLA_2020,Gelato2024,MPC2025,SODA2024}, and learning-based schemes like GreenABR+~\cite{GreenABR+2024}. These approaches primarily target human perceptual quality under bandwidth variation. However, VCM~\cite{VCM_2020} points out that machine-oriented transmission should prioritize task-relevant information over pixel reconstruction. Consequently, feature-based visual IoT transmission~\cite{Wang2019} delivers compact task-specific representations, while multi-stream offloading~\cite{hudieli} and AITransfer~\cite{AITransfer2023} incorporate content importance and network dynamics into adaptation. Nonetheless, a unified transmission solution that simultaneously supports structured semantic information and preserves rich visual evidence for downstream anomaly analysis is still lacking.

\section{Methods}

\subsection{Overall Design}

\begin{figure*}[t]
  \centering
  \includegraphics[width=0.9\textwidth]{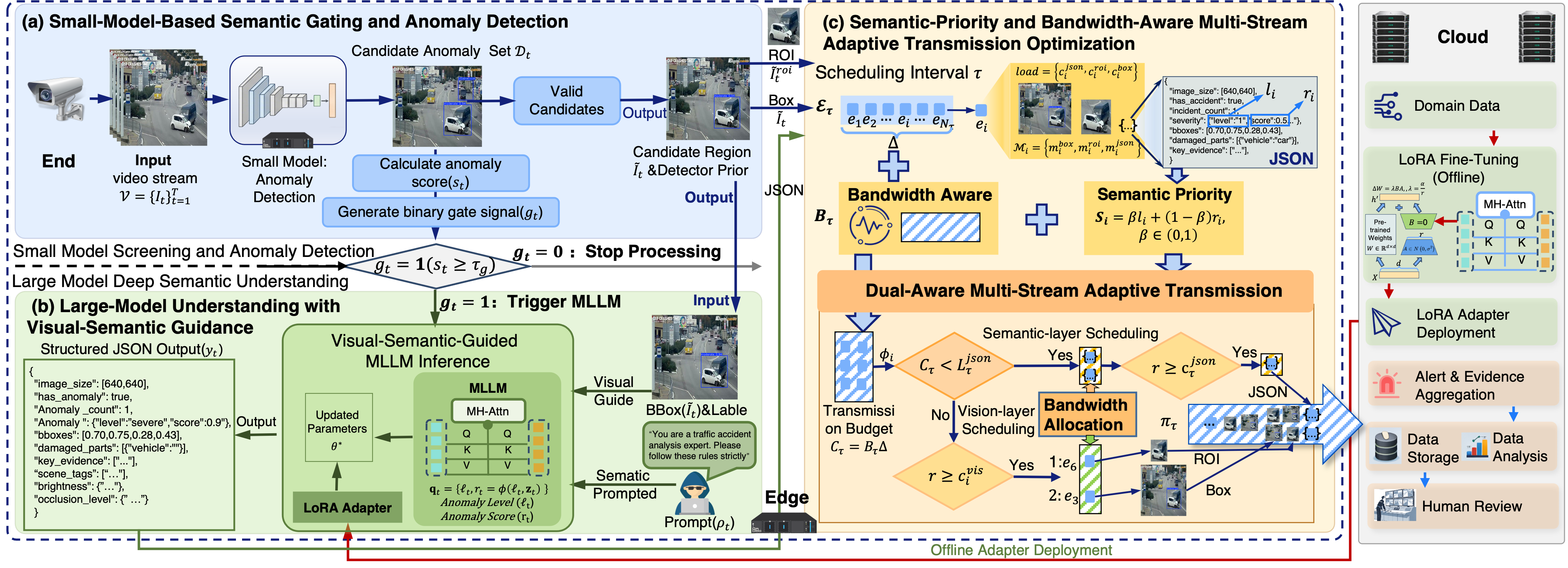}
  \caption{Overall workflow of DAT for dual-aware multi-stream transmission in edge--cloud anomaly detection.}
  \label{fig:DAT}
\end{figure*}

The overall design of DAT is as shown in Figure~\ref{fig:DAT}. End cameras continuously capture visual inputs and forward the video stream to nearby edge servers. Each edge server performs online anomaly gating, semantic anomaly alarm generation, and multi-stream transmission scheduling. The cloud server handles offline LoRA-based domain adaptation, alarm reception, evidence storage, and downstream verification. LoRA adapters~\cite{LoRA2021} are trained on the cloud server and distributed to edge servers. During online deployment, each edge server runs the lightweight frontend model and the adapter-equipped MLLM for inference~\cite{LLaMA-Adapter2023}; no large-model training is conducted locally. DAT first converts continuous visual inputs into two complementary streams at the edge and then schedules them according to the network state.  The semantic stream carries compact machine-readable anomaly alarms for timely detection, while the visual stream carries image-level evidence for later verification. DAT has two coupled modules. The small--large model collaborative cascaded inference module uses a lightweight frontend to filter candidate anomaly events and triggers the edge-side MLLM only for gated candidates. The MLLM uses frontend-provided spatial priors and task prompts to generate semantic anomaly alarms, confining expensive multimodal reasoning to high-value content. The Dual-Aware multi-stream transmission module schedules semantic and visual evidence streams according to event priority and time-varying uplink bandwidth: it prioritizes timely delivery of semantic anomaly alarms under congestion and progressively backfills visual evidence when residual bandwidth is available.

\subsection{Small--Large Cascaded Structured Semantic Generation}
\label{sec:cascaded_alert}

As shown in Figure~\ref{fig:DAT}(a)(b), DAT uses a cascaded small--large model design to avoid invoking the MLLM on every frame. Given an input video stream $\mathcal{V} =\{I_t\}_{t=1}^{T}$, the lightweight frontend model $f_s(\cdot)$ runs continuously on the edge server and produces candidate detections
  \begin{equation}
  D_t=f_s(I_t)=\{(b_{t,i},c_{t,i},p_{t,i})\}_{i=1}^{N_t},
  \end{equation}
where $b_{t,i}$, $c_{t,i}$, and $p_{t,i}$ denote the bounding box, category label, and confidence score of the $i$-th candidate, respectively. DAT defines the frame-level anomaly score as $s_t=\max_i p_{t,i}$ and triggers the MLLM only when $g_t=\mathbf{1}(s_t\ge\tau_g)$. For triggered frames, detections satisfying $p_{t,i}\ge\tau_g$ are retained as valid candidates. DAT uses boxed-image overlays $\tilde{I}_t$ as spatial priors, while both boxed images and ROI crops $\tilde{I}_t^{\mathrm{roi}}$ serve as visual evidence.

The adapter-equipped MLLM takes the guided visual input $\tilde{I}_t$ and task prompt $\rho_t$ as input and generates a structured semantic output:
  \begin{equation}
  y_t=f_l(\tilde{I}_t,\rho_t;\theta+\Delta\theta),
  \label{eq:mllm_inference}
  \end{equation}
where $f_l(\cdot)$ denotes the MLLM, $\theta$ denotes the frozen pretrained parameters, and $\Delta\theta$ denotes the LoRA adapter parameters. The output $y_t$ is represented as a machine-readable JSON object and forms the structured semantic stream  used by the downstream scheduler. To connect semantic generation with transmission scheduling, DAT extracts an event-level priority from $y_t$. For each candidate event $e_i$, DAT normalizes its discrete priority level $l_i\in\{1,\dots,L\}$ and  combines it with the continuous risk score $r_i\in[0,1]$:
  \begin{equation}
  \bar{l}_i=\frac{l_i-1}{L-1}, \quad
  S_i=\beta \bar{l}_i+(1-\beta)r_i,\quad \beta\in(0,1),
  \label{eq:event_priority}
  \end{equation}
where $S_i$ denotes the semantic priority score for transmission scheduling, and $\beta$ balances the normalized coarse priority and fine-grained risk. The structured output, priority score, and associated spatial priors are then passed to the  Dual-Aware transmission module.

\subsection{Semantic-Priority and Bandwidth-Aware Multi-Stream Transmission}

\subsubsection{Problem Formulation.}

As shown in Figure~\ref{fig:DAT}(c), after cascaded semantic-anomaly-alarm generation, the transmission scheduler operates only on candidate events that have passed frontend gating and have generated anomaly alarms. Let $\tau$ denote a scheduling interval and let $\mathcal{E}_{\tau}=\{e_i\mid i=1,\dots,N_{\tau}\}$ be the set of pending candidate events in this interval. Each event $e_i$ has a semantic priority score $S_i$, obtained from the structured output in Eq.~\ref{eq:event_priority}. For each event, DAT generates one structured semantic unit and two visual evidence candidates. In our implementation, the candidate transmission set is $M_i=\{m_i^{\mathrm{json}},m_i^{\mathrm{roi}},m_i^{\mathrm{box}}\}$ where $m_i^{\mathrm{json}}$ is the JSON alert, $m_i^{\mathrm{roi}}$ is the ROI evidence, and $m_i^{\mathrm{box}}$ is the boxed-image evidence. Their transmission costs are denoted by $c_i^{\mathrm{json}}$, $c_i^{\mathrm{roi}}$, and $c_i^{\mathrm{box}}$, respectively. Given the average available uplink bandwidth $B_{\tau}$ and interval duration $\Delta$, the transmission budget of interval $\tau$ is
$
  C_{\tau}=B_{\tau}\Delta 
$. We characterize this semantic-first objective through the following lexicographic formulation~\cite{lai2023pure,ehrgott2005multicriteria}:

  \begingroup
  \small
  \begin{equation}
  \label{eq:dual_aware_opt}
  \begin{aligned}[c]
  \underset{\{x_{i,\tau}^{k}\},\,\pi_{\tau},\,\{z_{i,\tau}^{\mathrm{vis}}\}}{\operatorname{lexmin}}
  &\left(
  \sum_{e_i\in\mathcal{E}_{\tau}} S_i \bar{T}_{i,\tau}^{\mathrm{sem}},
  -\sum_{e_i\in\mathcal{E}_{\tau}} S_i z_{i,\tau}^{\mathrm{vis}}
  \right) \\
  \mathrm{s.t.}\quad
  & \bar{T}_{i,\tau}^{\mathrm{sem}}
  =
  x_{i,\tau}^{\mathrm{json}}T_{i,\tau}^{\mathrm{sem}}
  +
  (1-x_{i,\tau}^{\mathrm{json}})T_{\mathrm{miss}},
   \forall e_i\in\mathcal{E}_{\tau},\\
  & \sum_{e_i\in\mathcal{E}_{\tau}}
  \sum_{k\in\{\mathrm{json},\mathrm{roi},\mathrm{box}\}}
  c_i^{k}x_{i,\tau}^{k}\le C_{\tau},\\
  & x_{i,\tau}^{\mathrm{roi}}\le x_{i,\tau}^{\mathrm{json}},\quad
  x_{i,\tau}^{\mathrm{box}}\le x_{i,\tau}^{\mathrm{json}},
  \quad \forall e_i\in\mathcal{E}_{\tau},\\
  & x_{i,\tau}^{\mathrm{roi}}+x_{i,\tau}^{\mathrm{box}}\le 1,
  \quad \forall e_i\in\mathcal{E}_{\tau},\\
  & z_{i,\tau}^{\mathrm{vis}}\le
  x_{i,\tau}^{\mathrm{roi}}+x_{i,\tau}^{\mathrm{box}},
  \quad \forall e_i\in\mathcal{E}_{\tau},\\
  & T_{i,\tau}^{\mathrm{vis}}\le D_{\mathrm{vis}}
  +M(1-z_{i,\tau}^{\mathrm{vis}}),
  \quad \forall e_i\in\mathcal{E}_{\tau},\\
  & \pi_{\tau}\in\Pi(U_{\tau}),\quad
  U_{\tau}=\{(i,k)\mid x_{i,\tau}^{k}=1\},\\
  & x_{i,\tau}^{k}\in\{0,1\},\quad
  z_{i,\tau}^{\mathrm{vis}}\in\{0,1\}.
  \end{aligned}
  \end{equation}
  \endgroup

DAT adopts a hierarchical transmission objective: it prioritizes semantic anomaly alarms for timely delivery and transmits visual evidence only when it does not delay semantic-anomaly-alarm delivery. Let $x_{i, \tau}^{k}\in\{0,1\}$ indicate whether unit $k\in\{\mathrm{json},\mathrm{roi},\mathrm{box}\}$ of event $e_i$ is selected in interval $\tau$, where $\mathrm{json}$ denotes the semantic-anomaly-alarm payload, and let $\pi_{\tau}$ denote the transmission order. $T_{i,\tau}^{\mathrm{sem}}$ is the completion time of the semantic anomaly alarm. Unselected semantic anomaly alarms incur a penalty $T_{\mathrm{miss}}$, which is larger than any feasible semantic completion time. Let $z_{i,\tau}^{\mathrm{vis}}\in\{0,1\}$ indicate whether the visual evidence of $e_i$ meets deadline $D_{\mathrm{vis}}$. The resulting lexicographic objective first minimizes weighted semantic-anomaly-alarm latency and then maximizes on-time visual delivery, subject to penalized semantic time, per-interval bandwidth, semantic-before-visual ordering, single-visual-unit rule, deadline validity, ordering validity, and binary decisions. Given this combinatorial structure, DAT adopts an online hierarchical greedy scheduler.

  \begin{algorithm}[th]
  \small
  \caption{Online hierarchical greedy scheduling for dual-aware transmission}
  \label{alg:elastic_tx}
  \textbf{Input}: $\mathcal{E}_{\tau}$, $\{S_i\}$, $
  \{c_i^{\mathrm{json}},c_i^{\mathrm{roi}},c_i^{\mathrm{box}}\}$, $B_\tau$, $\Delta$, $D_{\mathrm{vis}}$\\
  \textbf{Output}: $\{x_{i,\tau}^k\}$, $\pi_\tau$, $t$
  \begin{algorithmic}[1]
  \STATE $C_\tau \gets B_\tau \Delta$, $r \gets C_\tau$, $t \gets 0$, $\pi_\tau \gets \emptyset$
  \STATE $x_{i,\tau}^k \gets 0,\ \forall e_i\in\mathcal{E}_{\tau},\ k\in\{\mathrm{json},\mathrm{roi},
  \mathrm{box}\}$

  \FOR{each $e_i\in\mathcal{E}_{\tau}$}
      \STATE $\phi_i \gets S_i / c_i^{\mathrm{json}}$
  \ENDFOR
  \STATE Sort $\mathcal{E}_{\tau}$ in descending order of $\phi_i$

  \FOR{each $e_i\in\mathcal{E}_{\tau}$}
      \IF{$r \ge c_i^{\mathrm{json}}$}
          \STATE $x_{i,\tau}^{\mathrm{json}}\gets 1$, append $(i,\mathrm{json})$ to $\pi_\tau$
          \STATE $r \gets r-c_i^{\mathrm{json}}$, $t \gets t+c_i^{\mathrm{json}}/B_\tau$
      \ENDIF
  \ENDFOR

  \FOR{each $e_i$ with $x_{i,\tau}^{\mathrm{json}}=1$}
      \STATE $c_i^{\mathrm{vis}}\gets \min(c_i^{\mathrm{roi}},c_i^{\mathrm{box}})$, $k_i^*\gets \arg\min_{k\in\{\mathrm{roi},\mathrm{box}\}} c_i^k$
      \STATE $\psi_i \gets S_i / c_i^{\mathrm{vis}}$
  \ENDFOR
  \STATE Sort selected events in descending order of $\psi_i$

  \FOR{each selected $e_i$}
      \IF{$r \ge c_i^{\mathrm{vis}}$ \textbf{and} $t + c_i^{\mathrm{vis}}/B_\tau \le D_{\mathrm{vis}}$}
          \STATE $x_{i,\tau}^{k_i^*}\gets 1$, append $(i,k_i^*)$ to $\pi_\tau$
          \STATE $r \gets r-c_i^{\mathrm{vis}}$, $t \gets t+c_i^{\mathrm{vis}}/B_\tau$
      \ENDIF
  \ENDFOR

  \STATE \textbf{return} $\{x_{i,\tau}^k\}, \pi_\tau, t$
  \end{algorithmic}
  \end{algorithm}

\subsubsection{Online Hierarchical Greedy Scheduler.}

As shown in Algorithm~\ref{alg:elastic_tx}, DAT follows the lexicographic structure of Eq.~\eqref{eq:dual_aware_opt}: it schedules the structured semantic stream first and then uses the residual bandwidth to backfill the visual evidence stream. The scheduler takes as input the pending event set $\mathcal{E}_{\tau}$, event priorities $\{S_i\}$, transmission costs $\{c_i^{\mathrm{json}},c_i^{\mathrm{roi}},c_i^{\mathrm{box}}\}$, available bandwidth $B_{\tau}$, interval duration $\Delta$, and visual evidence deadline $D_{\mathrm{vis}}$. It first computes the interval budget $C_{\tau}=B_{\tau}\Delta$ and initializes the residual budget $r=C_{\tau}$, accumulated transmission time $t=0$, and transmission sequence $\pi_{\tau}=\emptyset$.

In the semantic layer, DAT assigns each event a semantic scheduling score $\phi_i=S_i/c_i^{\mathrm{json}}$, which measures the semantic-priority gain per unit JSON transmission cost. The scheduler scans events in descending order of $\phi_i$ and appends their JSON units to the transmission sequence whenever the residual budget permits. This stage prioritizes high-value and lightweight semantic anomaly alarms, approximating the semantic objective in Eq.~\eqref{eq:dual_aware_opt}. After the semantic layer, the set of events whose JSON alerts have been selected is
  \begin{equation}
  \mathcal{E}_{\tau}^{\mathrm{json}}
  =\{e_i\in\mathcal{E}_{\tau}\mid x_{i,\tau}^{\mathrm{json}}=1\}.
  \end{equation}
  
In the visual layer, DAT considers only events in $\mathcal{E}_{\tau}^{\mathrm{json}}$, ensuring that visual evidence never precedes its corresponding semantic anomaly alarm. For each event with a selected JSON alert, the scheduler chooses the lower-cost visual unit from ROI and Box:
  \begin{equation}
  k_i^*=\arg\min_{k\in\{\mathrm{roi},\mathrm{box}\}}c_i^k,\qquad
  c_i^{\mathrm{vis}}=\min(c_i^{\mathrm{roi}},c_i^{\mathrm{box}}).
  \end{equation}
It then defines the visual scheduling score as
  \begin{equation}
  \psi_i=\frac{S_i}{c_i^{\mathrm{vis}}},\qquad e_i\in \mathcal{E}_{\tau}^{\mathrm{json}}.
  \end{equation}
The scheduler scans visual candidates in descending order of $\psi_i$ and appends a visual unit only when both the residual-budget and deadline conditions are satisfied:
  \begin{equation}
  r\ge c_i^{\mathrm{vis}},\qquad
  t+\frac{c_i^{\mathrm{vis}}}{B_{\tau}}\le D_{\mathrm{vis}}.
  \end{equation}

\subsubsection{Ordering Optimality and Complexity Analysis.}
For a fixed set of selected JSON units and constant bandwidth within interval $\tau$, ordering JSON alerts by non-increasing $\phi_i=S_i/c_i^{\mathrm{json}}$ minimizes their semantic-priority-weighted completion time. This follows from the adjacent-exchange argument: for two adjacent units $i$ and $j$, placing $i$ before $j$ is no worse than the reverse order if and only if \(S_i/c_i^{\mathrm{json}} \ge S_j/c_j^{\mathrm{json}}\). The result establishes the optimal order for a fixed JSON set, while selecting transmission units under a discrete bandwidth budget remains combinatorial and is approximated by Algorithm~\ref{alg:elastic_tx}. The algorithm also preserves  feasibility by checking the JSON-before-visual dependency, non-redundancy, residual-budget, visual-deadline, ordering, and binary-selection constraints before appending each unit. Its dominant cost is sorting $N_{\tau}$ candidate events in each interval, giving $O(N_{\tau}\log N_{\tau})$ time and $O(N_{\tau})$ space complexity for online execution on edge servers.

\begin{table*}[t]
  \centering
  \setlength{\tabcolsep}{4.5pt}
  \renewcommand{\arraystretch}{1.08}
  \small
  \begin{tabular}{@{}lccccccccccc@{}}
    \toprule
    \multirow{3}{*}{Method}
    & \multicolumn{4}{c}{Anomaly Detection}
    & \multicolumn{3}{c}{Structured Output}
    & \multicolumn{4}{c}{System Cost / Load} \\
    \cmidrule(lr){2-5}
    \cmidrule(lr){6-8}
    \cmidrule(l){9-12}

    & Recall
    & Precision
    & F1
    & FPR
    & Count EM
    & Sev. F1
    & BBox mIoU
    & Alert-Path
    & MLLM
    & Visual Token
    & Upload \\

    & (\%)$\uparrow$
    & (\%)$\uparrow$
    & (\%)$\uparrow$
    & (\%)$\downarrow$
    & (\%)$\uparrow$
    & (\%)$\uparrow$
    & (\%)$\uparrow$
    & P95 (s)$\downarrow$
    & Calls (\%)$\downarrow$
    & (M)$\downarrow$
    & (MB)$\downarrow$ \\
    \midrule

    Edge-Only
    & \underline{66.25}
    & 92.54
    & \underline{77.22}
    & 25.76
    & \underline{59.30}
    & 29.93
    & --
    & 6.322
    & 100.00
    & 1.44
    & -- \\

    Cloud-Only
    & 53.85
    & \underline{98.85}
    & 69.72
    & \textbf{3.03}
    & 59.17
    & \underline{74.53}
    & \underline{23.73}
    & \textbf{3.932}
    & 100.00
    & 1.44
    & 54.73 \\

    SAEC
    & 49.45
    & 98.75
    & 65.90
    & \textbf{3.03}
    & 55.53
    & 71.85
    & 21.80
    & 4.824
    & 88.95
    & 1.28
    & 48.68 \\

    EdgeCloudAI
    & 23.70
    & 94.97
    & 37.94
    & 6.06
    & 34.59
    & 52.38
    & 7.91
    & 5.101
    & \textbf{72.04}
    & \textbf{0.87}
    & \underline{38.04} \\

    \textbf{DAT}
    & \textbf{96.39}
    & \textbf{99.19}
    & \textbf{97.77}
    & \underline{3.79}
    & \textbf{92.98}
    & \textbf{93.31}
    & \textbf{69.49}
    & \underline{4.771}
    & \underline{80.49}
    & \underline{1.16}
    & \textbf{23.91} \\

    \bottomrule
  \end{tabular}

  \caption{Edge--cloud deployment comparison of DAT and representative baselines. The best and second-best results are highlighted in bold and underlined, respectively.}
  \label{tab:alarm_structured_cost}
\end{table*}

\section{Experiments}

\subsection{Evaluation Setup}

\subsubsection{Datasets and Implementation Details.}

We adopt traffic-accident detection as the main evaluation setting and choose the Accidents Detection Dataset~\cite{accidents_detection_dataset} as the primary benchmark, with 10,469/1,004/649 images for training, validation, and testing. To improve non-accident discrimination, we add Non-Accident samples from Accident Detection From CCTV Footage~\cite{CCTV_Footage2020} as external negatives; after resizing and deduplication, 508 images are split into 268/120/120. We further construct an ACCIDENT~\cite{ACCIDENT2026} continuous-video workload with 1,520 accident videos, 72,638 sampled frames, and accident-time annotations for event-level evaluation. For cross-domain evaluation, we use IPAD~\cite{IPAD2024}, which covers 16 types of industrial equipment, with 48/32/212 video sequences for LoRA adaptation, validation, and testing. Experiments run on an NVIDIA RTX 5090 32GB edge server and an NVIDIA A100 40GB cloud server. DAT deploys YOLOv12s~\cite{2025yolov12} and Qwen2.5-VL-7B-Instruct~\cite{2025qwen2.5vl} on the edge server for online inference. LoRA fine-tuning is performed on the cloud server with Qwen2.5-VL-7B-Instruct, and the resulting adapters~\cite{LoRA2021} are distributed to the edge server. Transmission evaluation uses the Zoom1 uplink trace from the 5G Traffic Datasets~\cite{5G_Traffic_dataset}. Upload volume is measured only on the edge-to-cloud link and excludes local camera-to-edge ingestion. Bandwidth measurement and scheduling use 1~s intervals. 



\subsubsection{Parameter Selection.}
\begin{figure}[h]
  \centering
  \includegraphics[width=\linewidth]{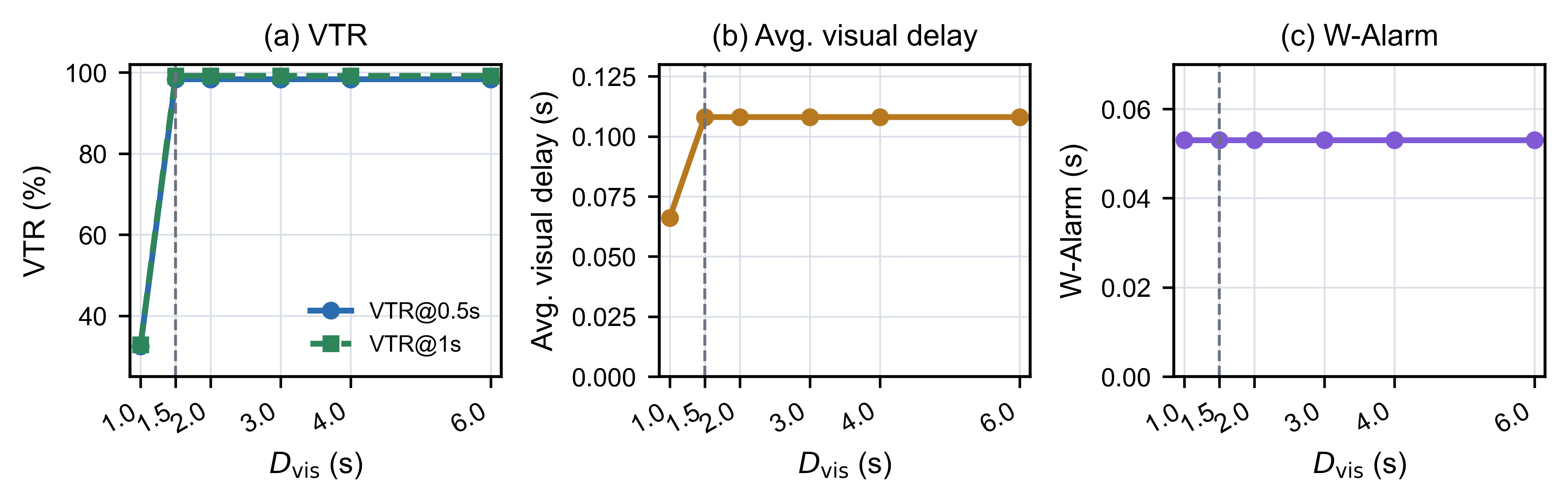}
  \caption{Parameter analysis of $D_{vis}$.}
  \label{fig:dvis}
\end{figure}

We test $\{\mathrm{low},\,\mathrm{medium},\,\mathrm{burst}\}$ arrival patterns and $\{1.0\times,0.5\times,0.25\times\}$ bandwidth scales, with $D_{\mathrm{vis}}=1.5$~s by default. DAT keeps one representative ROI per traffic event, selected by severity and then data size. All scheduling methods use the same mixed stream of 600 accident and 600 non-accident samples, bandwidth traces, arrival patterns, and scheduling  intervals. Figure~\ref{fig:dvis} analyzes $D_{\mathrm{vis}}$ under burst arrivals and $0.25\times$ bandwidth. Increasing $D_{\mathrm{vis}}$ from 1.0~s to 1.5~s admits more visual evidence into the deadline-valid backfilling stage; beyond 1.5~s, visual-delivery metrics stabilize and the weighted anomaly-alarm latency (W-Alarm) stays near 0.053~s. We therefore set $D_{\mathrm{vis}}=1.5$~s.

\subsubsection{Baselines and Metrics.}
We compare DAT from two perspectives: system performance and multi-stream transmission performance. The system baselines include MobileVLM-V2-1.7B (Edge-Only)~\cite{mobilevlmv2_2024}, Qwen2.5-VL-7B (Cloud-Only)~\cite{2025qwen2.5vl}, SAEC~\cite{SAEC}, and EdgeCloudAI~\cite{EdgeCloudAI2024} (reimpl.), where SAEC performs scene-complexity-aware edge--cloud routing and EdgeCloudAI applies edge-side filtering before cloud-side MLLM inference. Edge-Only is evaluated only on the image benchmark; for continuous-video and cross-domain workloads, Cloud-Only serves as the full-MLLM reference. Transmission baselines include SODA~\cite{SODA2024} and MAOS~\cite{hudieli}. 

Metrics include anomaly detection performance (Recall, Precision, F1, and FPR), structured-output quality when annotations are available (Count EM, Severity F1, and BBox mIoU),  system cost (Alert-Path P95, First-Alert Latency, MLLM Call Rate, visual-token load, and Total Upload), and transmission timeliness (W-Alarm, VTR@0.5s/1s, and Avg Visual Delay). For continuous-video evaluation, we additionally report Event Recall@$\pm2$s on ACCIDENT and Event Recall@$\pm 16$f on IPAD.

\begin{figure}[h]
  \centering
  \includegraphics[width=\linewidth]{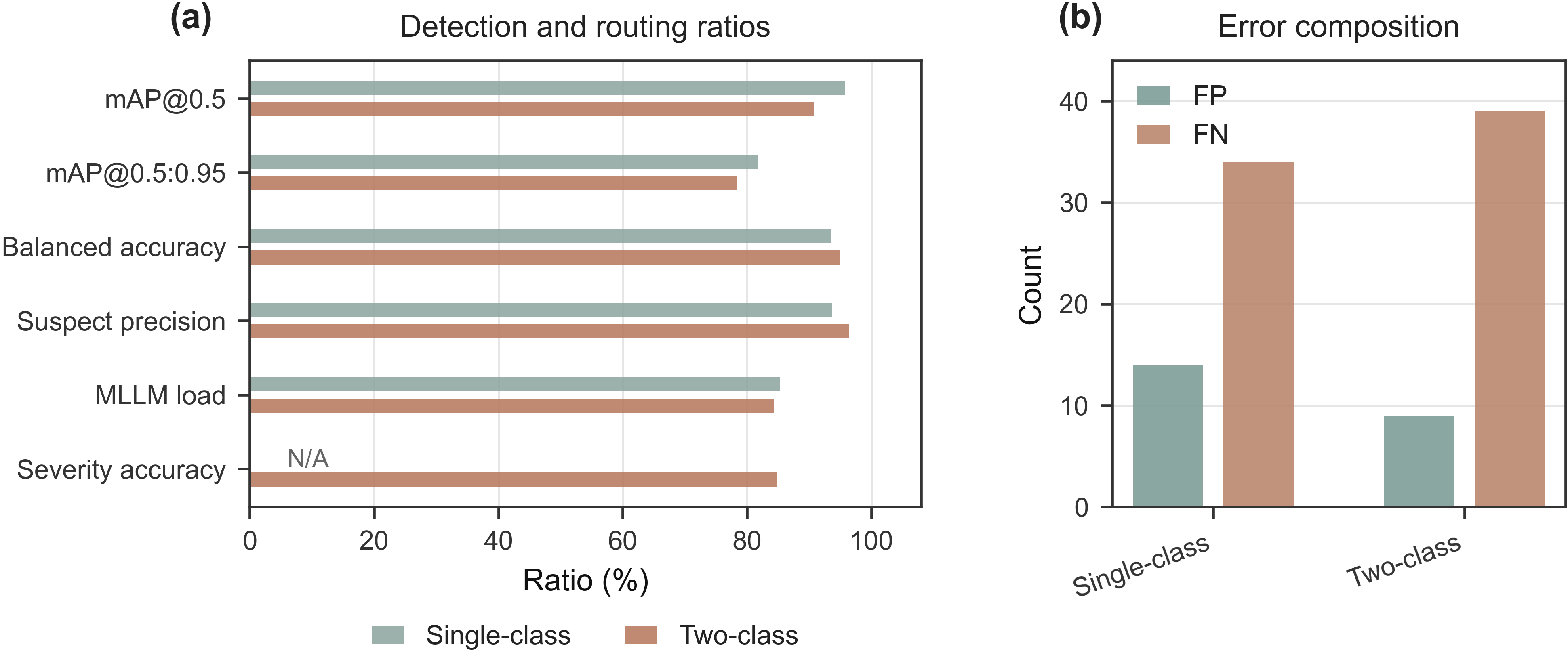}
  \caption{Small-model detector design for system-level gating.}
  \label{fig:small_model_gating}
\end{figure}

\subsection{Experimental Results}

\subsubsection{Edge--Cloud Deployment Performance Comparison.}
Table~\ref{tab:alarm_structured_cost} compares DAT with representative edge--cloud deployment schemes. DAT achieves 96.39\% Recall and 97.77\% F1, and consistently ranks first across anomaly detection and structured-output metrics. Edge-Only~\cite{mobilevlmv2_2024} obtains relatively high Recall and F1, but suffers from a high FPR (25.76\%) and weak severity prediction (29.93\% Sev. F1). Cloud-Only~\cite{2025qwen2.5vl} and SAEC~\cite{SAEC} keep FPR low, but both achieve below 54\% Recall, while EdgeCloudAI~\cite{EdgeCloudAI2024} reduces load more aggressively at the cost of a much lower F1 (37.94\%). DAT provides a stronger accuracy--cost balance. It reduces total upload to 23.91 MB, which is 56.3\%, 50.9\%, and 37.1\% lower than Cloud-Only, SAEC, and EdgeCloudAI, respectively. Its visual-token load is 1.16M, reducing token usage by 9.4\%--19.4\% compared with Cloud-Only and SAEC.  Overall, DAT trades a small increase in tail latency over Cloud-Only for substantially higher accident recall, stronger structured output, and the lowest upload volume.

Figure~\ref{fig:small_model_gating} further explains the source of this system-level cost reduction. We use $\tau_{\mathrm{low}}=0.25$ as the base threshold and introduce $\tau_{\mathrm{high}}=0.8$ to compare a single-class accident detector with a two-class moderate/severe detector. Although the single-class setting gives slightly higher mAP, the two-class setting improves Balanced Accuracy, Suspect Precision, False Positive control, and MLLM Load, while also providing severity priors for cascaded inference. We therefore use the two-class lightweight frontend as the default gating configuration in the following experiments.

  \begin{table}[th]
  \centering
  \small
  \setlength{\tabcolsep}{4pt} 
  \begin{tabular}{lcccc}
    \toprule
    Method & \makecell{W-Alarm\\(s)} & \makecell{VTR@0.5s\\(\%)} & \makecell{VTR@1s\\(\%)} & \makecell{Avg\\Visual\\Delay\\(s)} \\
    \midrule
    \rowcolor{gray!15}
    \multicolumn{5}{l}{Block A. Medium (0.25$\times$)} \\
    SODA  & 0.1253 & 99.33 & 99.83 & 0.1212 \\
    MAOS  & 0.1530 & 73.67 & 73.67 & 0.1972 \\
    Bandwidth-Only & 0.0543 & 94.67 & 95.17  & 0.1742 \\
    Priority-Only  & 0.0509 & \textbf{99.83} & 99.83 & 0.0637 \\
    \textbf{DAT}    & \textbf{0.0508} & \textbf{99.83} & \textbf{100.0} & \textbf{0.0368} \\
    \midrule
    \rowcolor{gray!15}
    \multicolumn{5}{l}{Block B. Burst (0.25$\times$)} \\
    SODA  & 0.4381 & 65.50 & 97.50 & 0.4324 \\
    MAOS  & 0.3041 & 56.83 & 70.83 & 0.3547 \\
    Bandwidth-Only & 0.2361 & 80.17 & 89.17  & 0.4392 \\
    Priority-Only  & 0.1704 & 92.33 & 98.00 & 0.2153 \\

    \textbf{DAT}    & \textbf{0.0531} & \textbf{98.33} & \textbf{99.17} & \textbf{0.1081} \\
    \midrule
    \rowcolor{gray!15}
    \multicolumn{5}{l}{Block C. Alert-Carrier Design Study under Burst (0.25$\times$)} \\
    JSON-Only        & \textbf{0.0531} & 0.00  & 0.00  & - \\
    Fixed Box Upload & 0.6338 & 82.00 & 96.33 & 0.3436 \\
    Fixed ROI Upload & 0.6512 & 77.50 & 92.83 & 0.3382 \\
    Fixed JSON+Box   & 0.2602 & 81.83 & 96.33 & 0.3460 \\
    \textbf{DAT}      & \textbf{0.0531} & \textbf{98.33} & \textbf{99.17} & \textbf{0.1081} \\
    \bottomrule
  \end{tabular}
  \caption{Overall performance of semantic-priority- and bandwidth-aware multi-stream adaptive transmission.}
  \label{tab:overall_performance}
\end{table}

\subsubsection{Dual-Aware Multi-Stream Transmission Performance.}

Table~\ref{tab:overall_performance} demonstrates the advantage of DAT for multi-stream scheduling under dynamic bandwidth. Overall, DAT achieves the lowest W-Alarm in both medium and burst arrivals under $0.25\times$ bandwidth, while maintaining the highest or near-highest VTR. In the more challenging Block B setting (burst  arrivals, $0.25\times$ bandwidth), DAT reduces W-Alarm to 0.0531 s, corresponding to reductions of 87.9\%, 82.5\%, 77.5\%, and 68.8\% over SODA, MAOS, Bandwidth-Only, and Priority-Only, respectively. It also improves VTR@0.5s to 98.33\% and reduces Avg Visual Delay to 0.1081 s. These results show that DAT coordinates semantic anomaly alarms and visual evidence backfilling more effectively under concurrent arrivals and severe bandwidth constraints. Block C further shows that JSON-Only minimizes semantic-anomaly-alarm latency but removes visual verifiability, whereas fixed Box, ROI, and JSON+Box transmission incur larger alarm and visual delays. DAT resolves this tension by transmitting compact JSON alerts first and allocating residual bandwidth to visual evidence, thereby preserving both alert timeliness and evidence availability under congestion.

\begin{figure}[h]
  \centering
  \includegraphics[width=\linewidth]{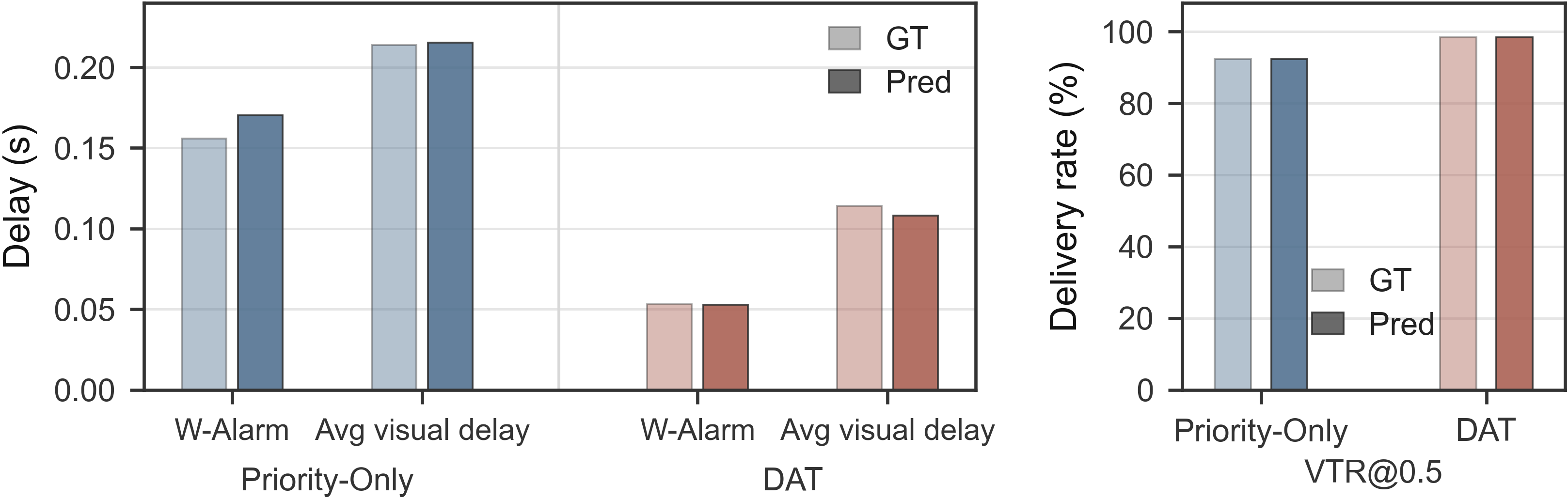}
  \caption{Validation under different priority sources.}
  \label{fig:priority_reliability}
\end{figure}

\subsubsection{Priority Prediction Reliability.}

Figure~\ref{fig:priority_reliability} compares DAT under ground-truth and predicted priorities. Under burst arrivals and $0.25\times$ bandwidth, DAT remains close to the ground-truth setting in W-Alarm, VTR@0.5s, and Avg Visual Delay, and consistently outperforms Priority-Only. This indicates that DAT's gain comes primarily from  jointly scheduling semantic priority and bandwidth dynamics, rather than from assuming perfect priority labels.

\begin{figure}[h]
  \centering
  \includegraphics[width=\linewidth]{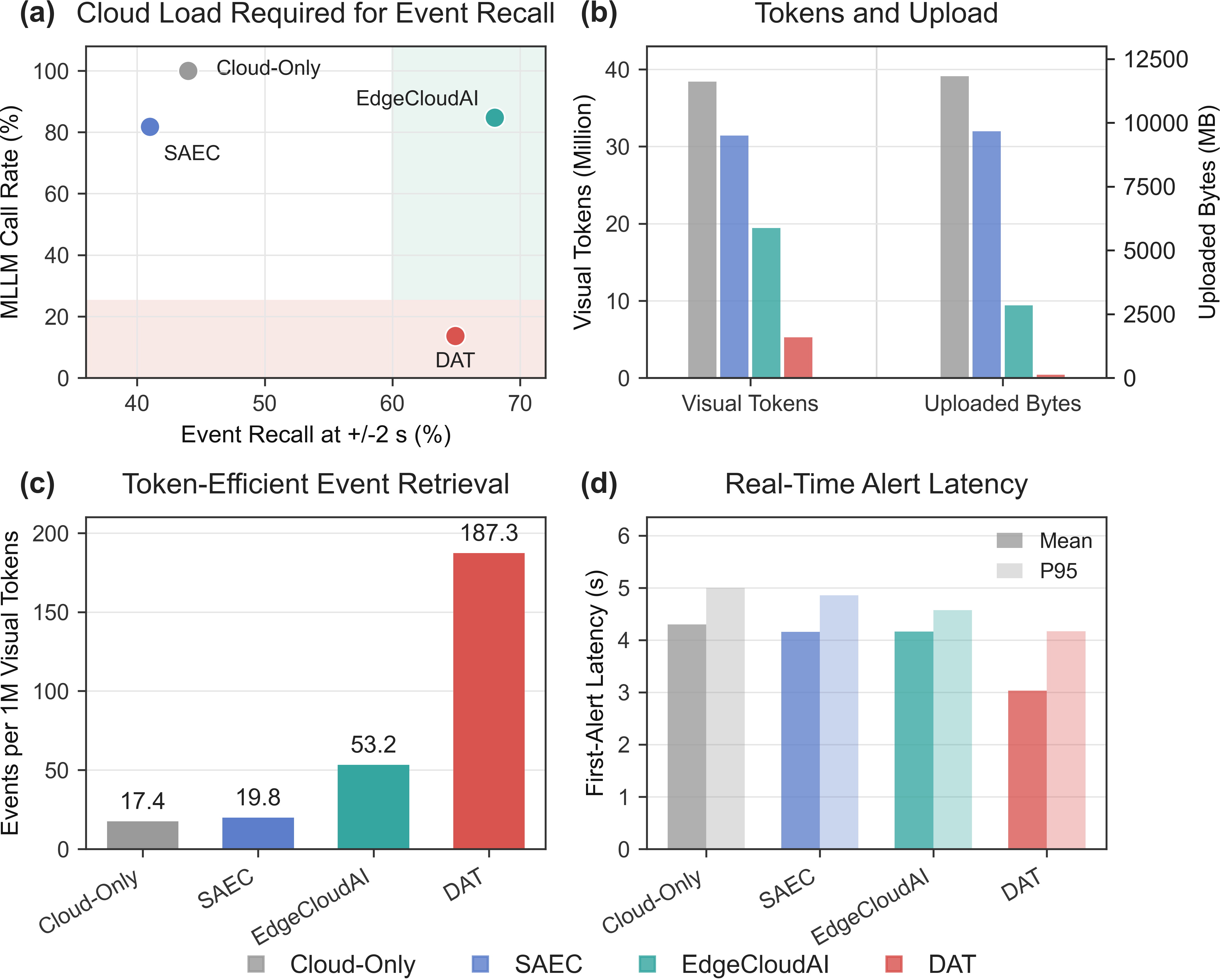}
  \caption{Continuous-video system evaluation on ACCIDENT.}
  \label{fig:continuous_video}
\end{figure}

\begin{table*}[!t]
  \centering
  \small
  \setlength{\tabcolsep}{5pt}
  \renewcommand{\arraystretch}{1.08}
  \begin{tabular}{lllcccccc}
    \toprule
    MLLMs & Input & Tuning & Count EM$\uparrow$ & Count MAE$\downarrow$ & SevAcc$\uparrow$ & SevF1$\uparrow$ & BBox mIoU$\uparrow$ & Recall@0.5$\uparrow$ \\
    \midrule
    Qwen2.5-VL-7B-Instruct & Raw image   & None & 78.98\% & 0.2150 & 75.36\% & 70.01\% & 16.71\% & 0.70\% \\
    Qwen2.5-VL-7B-Instruct & Raw image   & LoRA & 80.86\% & 0.1943 & 84.47\% & 83.62\% & 54.40\% & 68.71\% \\
    GPT-5 & Boxed image & None & \textbf{96.30\%} & \textbf{0.0370} & 98.49\% & 98.36\% & 53.12\% & 58.09\% \\
    \textbf{DAT} & Boxed image & LoRA & 95.50\% & 0.0500 & \textbf{98.83\%} & \textbf{98.74\%} & \textbf{68.33\%} & \textbf{84.83\%} \\
    \bottomrule
  \end{tabular}
  \caption{Evaluation of input guidance and domain adaptation for structured accident understanding.}
  \label{tab:structured_accident}
\end{table*}

\subsubsection{Continuous-Video Evaluation on ACCIDENT.}

Figure~\ref{fig:continuous_video} reports the system-level results on the ACCIDENT~\cite{ACCIDENT2026} continuous-video workload. DAT achieves 64.93\% Event Recall@$\pm2$s with an MLLM call rate of only 13.71\%, showing competitive accident-anomaly detection with substantially fewer large-model invocations. Although EdgeCloudAI obtains slightly higher recall, it requires about 6.18 times more MLLM calls than DAT, indicating that its recall gain comes at a substantially higher large-model invocation cost. DAT reduces the visual-token load to 5.27M and the upload volume to 121.85 MB; compared with Cloud-Only, SAEC, and  EdgeCloudAI, this corresponds to 72.86\%--86.29\% lower visual-token load and 95.72\%--98.97\% lower upload volume. DAT retrieves 187.31 events per million visual tokens, 3.52 times higher than the best baseline, and reduces the mean/P95 first-alert latency from 4.298/4.999 s for Cloud-Only to 3.030/4.168 s. Together, these results show that continuous video magnifies the redundancy of frame-level visual reasoning and upload, whereas DAT preserves event-level recall by converting dense frame streams into sparse semantic anomaly alarms and supplementing only lightweight visual evidence.

\begin{figure}[h]
  \centering
  \includegraphics[width=\linewidth]{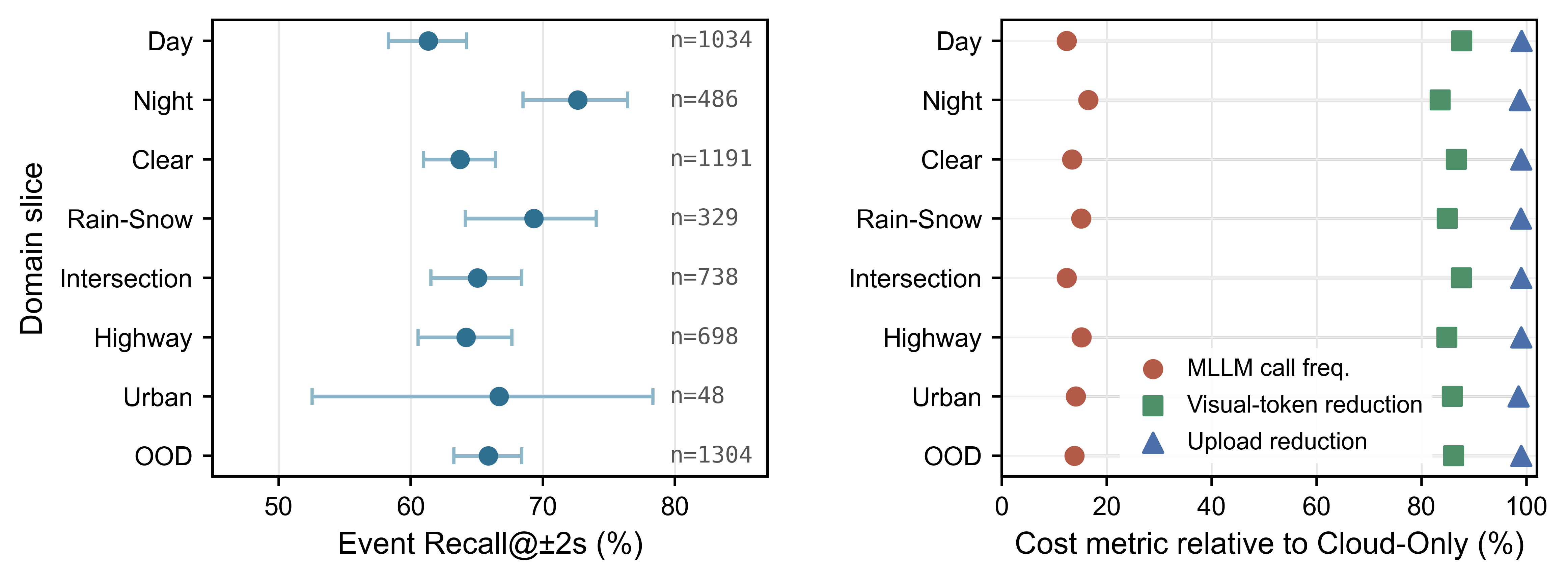}
  \caption{Domain-stratified robustness of DAT on ACCIDENT videos.}
  \label{fig:fanhua}
\end{figure}

\subsubsection{Domain-Stratified Robustness Analysis.}

We further examine DAT's robustness across common video-domain shifts. Using the official scene and environment tags, we construct eight ACCIDENT slices: Day, Night, Clear, Rain-Snow, Intersection, Highway, Urban, and OOD. As shown in Figure~\ref{fig:fanhua}, DAT maintains 61.32\%--72.63\% Event Recall@$\pm2$s across these slices, while keeping the MLLM call rate at only 12.35\%--16.46\% of Cloud-Only and reducing visual-token load and upload volume by 83.54\%--87.65\% and 98.51\%--99.08\%, respectively. These results show that DAT's event-level filtering remains effective under variations in lighting, weather, road type, and OOD distribution.

\subsubsection{Ablation and Further Analysis.}
Table~\ref{tab:structured_accident} shows that DAT's main advantage lies in severity recognition and spatial localization. LoRA domain adaptation substantially improves the raw-image model on these two aspects, increasing Severity F1 from 70.01\% to 83.62\% and BBox mIoU from 16.71\% to 54.40\%. Combining LoRA adaptation with boxed-image guidance, DAT achieves the  best localization-aware structured understanding, reaching 98.74\% Severity F1 and 68.33\% BBox mIoU, both higher than GPT-5. On the same 600 boxed-image subset, DAT also achieves 100\% Parse, Schema, and Rule compliance with zero violations, indicating stable structured-format generation.

Table~\ref{tab:IPAD} presents the cross-domain results on IPAD~\cite{IPAD2024} industrial anomaly videos. DAT achieves 35.62\% Event Recall@$\pm 16$f, about 3.06 times the strongest baseline, and also obtains the highest F1. This indicates that the DAT pipeline can transfer from traffic-accident detection to industrial anomaly streams with lightweight domain adaptation. DAT reduces the MLLM Call rate to 25.95\%, while lowering visual-token load by 23.26\%--74.04\% and upload volume by 91.68\%--97.21\% compared with the other baselines. Although DAT has higher FPR and Alert-Path P95 than the best baselines, the results show a clear high-recall, low-load operating point for cross-domain anomaly detection.



\begin{table}[t]
  \centering
  \small
  \setlength{\tabcolsep}{3.5pt}
  \renewcommand{\arraystretch}{1.08}
  \begin{tabular}{@{}lcccc}
    \toprule
    Metric
    & \makecell{Cloud-\\Only}
    & SAEC
    & \makecell{EdgeCloud-\\AI}
    & \textbf{DAT} \\
    \midrule

    \rowcolor{gray!15}
    \multicolumn{5}{@{}l}{Anomaly Detection} \\

    Event Recall@$\pm 16$f (\%) $\uparrow$
    & \underline{11.64}
    & 10.96
    & \underline{11.64}
    & \textbf{35.62} \\

    Precision (\%) $\uparrow$
    & 9.88
    & 9.76
    & \underline{11.26}
    & \textbf{11.98} \\

    F1 (\%) $\uparrow$
    & 10.69
    & 10.32
    & \underline{11.45}
    & \textbf{17.93} \\

    FPR (\%) $\downarrow$
    & 3.20
    & \underline{3.05}
    & \textbf{2.63}
    & 5.44 \\

    \addlinespace[2pt]
    \rowcolor{gray!15}
    \multicolumn{5}{@{}l}{System Cost / Load} \\

    Alert-Path P95 (s) $\downarrow$
    & \textbf{2.12}
    & \underline{2.17}
    & 2.21
    & 3.65 \\

    MLLM Call (\%) $\downarrow$
    & 100.00
    & 88.07
    & \underline{43.95}
    & \textbf{25.95} \\

    Visual-Token (M) $\downarrow$
    & 15.24
    & 13.42
    & \underline{5.16}
    & \textbf{3.96} \\

    Upload (MB) $\downarrow$
    & 719.54
    & 650.15
    & \underline{240.85}
    & \textbf{20.04} \\

    \bottomrule
  \end{tabular}
  \caption{Cross-domain evaluation on continuous industrial anomaly streams from IPAD. }
  \label{tab:IPAD}
\end{table}



\section{Conclusion}
In this paper, we present DAT, an edge--cloud collaborative framework for bandwidth-constrained visual anomaly detection. DAT combines small--large model cascaded inference, spatial-prior-guided structured understanding, and semantic-priority- and bandwidth-aware scheduling to transform continuous video into compact anomaly alarms and selected visual evidence. Experiments on traffic-accident images, continuous videos, and industrial anomaly streams demonstrate improved anomaly understanding with fewer MLLM invocations and shorter weighted alarm delay. Cross-domain results further indicate that DAT can transfer to new anomaly scenarios through lightweight adaptation.


\bibliography{aaai2027}


\end{document}